\documentclass[sigconf]{acmart}
\AtBeginDocument{%
  \providecommand\BibTeX{{%
    \normalfont B\kern-0.5em{\scshape i\kern-0.25em b}\kern-0.8em\TeX}}}

\setcopyright{rightsretained}
\copyrightyear{2024}
\acmYear{2024}
\acmConference[Adjunct Proceedings CHI '24]{}{May 11--16,
  2024}{Honolulu, HI, USA}
\begin{document}

\title{Demystifying CO\textsubscript{2}: lessons from nutrition labeling and step counting}

\author{Alexandre Filipowicz}
\email{alex.filipowicz@tri.global}
\orcid{0000-0002-1311-386X}
\affiliation{%
  \institution{Toyota Research Institute}
  \streetaddress{4440 El Camino Real}
  \city{Los Altos}
  \state{California}
  \country{USA}
  \postcode{94022}
}

\author{David A. Shamma}
\email{ayman.shamma@tri.global}
\orcid{0000-0003-2399-9374}
\affiliation{%
  \institution{Toyota Research Institute}
  \streetaddress{4440 El Camino Real}
  \city{Los Altos}
  \state{California}
  \country{USA}
  \postcode{94022}
}

\author{Vikram Mohanty}
\orcid{}
\affiliation{
  \institution{Bosch Research North America}
  \city{Sunnyvale}
  \state{California}
  \country{USA}
}

\author{Candice L. Hogan}
\email{candice.hogan@tri.global}
\orcid{0000-0002-3240-2560}
\affiliation{%
  \institution{Toyota Research Institute}
  \streetaddress{4440 El Camino Real}
  \city{Los Altos}
  \state{California}
  \country{USA}
  \postcode{94022}
}

\renewcommand{\shortauthors}{Filipowicz, et al.}

\begin{abstract}
  There is growing concern about climate change and increased interest in taking action. However, people have difficulty understanding abstract units like CO\textsubscript{2} and the relative environmental impact of different behaviors. This position piece explores findings from nutritional labeling and step counting research, two domains aimed at making abstract concepts (i.e., calories and exercise) more familiar to the general public. Research in these two domains suggests that consistent, widespread communication can make people more familiar and think more precisely about abstract units, but that better communication and understanding does not guarantee behavior change. These findings suggest that consistent and ubiquitous communication can make CO\textsubscript{2} units more familiar to people, which in turn could help interventions aimed at encouraging more sustainable behaviors.
\end{abstract}






\maketitle

\section{Introduction}
People around the world are growing increasingly concerned about their impact on the climate. A 2024 international survey identified that 86\% of respondents agreed that people in their countries should fight climate change~\cite{andre2024globally}. People also want to take action -- 69\% of respondents in the same survey were willing to donate 1\% of their annual salary if it could help tackle climate change~\cite{andre2024globally}. However, knowing which behaviors to change and in what respect can be challenging. People have difficulty estimating the real climate impact of their behaviors, tending to overestimate the impact of relatively-low intensity behaviors (e.g., turning off lights) and underestimating the impact of higher-intensity behaviors (e.g., long flights)~\cite{attari2010public, nielsen2021psychology, wynes2018measuring}. This misunderstanding of the relative impact of different behaviors can thus lead individuals to focus their efforts on changing behaviors that do relatively little to mitigate their overall impact on the environment~\cite{nielsen2021psychology}.

To improve people's understanding of their impact on the environment, user interface designers have explored ways to improve carbon literacy~\cite{sanguinetti2018information, nielsen2021psychology, mohanty2023save, west2016evaluating, epa_calc_2022, mulrow2019state}. Carbon footprint calculators have been designed to help people see and track the environmental impact of different behaviors~\cite{west2016evaluating, epa_calc_2022, mulrow2019state}. Additionally, interface designers are increasingly interested in displaying carbon information like CO\textsubscript{2} at decision points to encourage people to make more sustainable choices. These types of interventions have shown some promise: displaying information about carbon emissions can successfully help people choose less carbon intensive flights~\cite{sanguinetti2022nudging}, ground transportation choices~\cite{mohanty2023save, filipowicz2023promoting} and items on a food menu~\cite{beyer2024does}. 

Although many of these efforts have shown that information about carbon can influence people's choices, this is not sufficient to improve people's carbon literacy. Mohanty and colleagues~\cite{mohanty2023save} found that although information about CO\textsubscript{2} helped people make more sustainable transportation choices, and people preferred information about CO\textsubscript{2} over many other equivalencies (e.g., trees or gallons of gasoline), metrics like CO\textsubscript{2} are too abstract and people expressed that they did not have a good understanding of how different quantities of CO\textsubscript{2} relate to climate change.

In this position piece we explore findings from two other domains, nutrition labeling and step counting, that encountered similar issues of trying to communicate abstract metrics (e.g., calories, physical activity) to the general public. We examine the aspects of these efforts that worked and those that failed and relate these to current efforts aimed at improving carbon literacy. From this work, we derive a set of recommendations about ways in which interfaces can help improve carbon literacy and lead people to make more consistent sustainable choices.

\section{Related work}
There are a number of systems that aim to educate and help people track the environmental impact of different behaviors. Carbon footprint calculators are a common interface used to help improve carbon literacy~\cite{west2016evaluating, epa_calc_2022, mulrow2019state}. These calculators aim to help people track the emission intensity of everyday behaviors (e.g., home energy usage, food consumption, and transportation), visualize the impact of these different behaviors, and in some cases provide behavioral recommendations to reduce people's impact on the environment~\cite{west2016evaluating, mulrow2019state}. However, carbon calculators can be difficult to use effectively. They can require considerable effort on the part of the user (e.g., frequent input), are not always comprehensive (e.g., focus on home energy but not transportation), and do not always provide the information people need to remain motivated~\cite{mulrow2019state}. Consequently, people using carbon calculators tend not to remember the insights these tools provide and lose motivation for their usage~\cite{mulrow2019state}.

Designers have also explored ways of embedding different forms of \emph{eco-feedback} into user interfaces to notify users of the actual or potential CO\textsubscript{2} emissions related to their behaviors, and encourage more sustainable choices~\cite{sanguinetti2018information, mohanty2023save, filipowicz2023promoting, sanguinetti2022nudging}. Sanguinetti and colleagues~\cite{sanguinetti2018information} offer a comprehensive framework to help designers identify factors needed for effective eco-feedback, including the properties of the information being presented, the information timing, and the type of display. Eco-feedback interfaces tend to focus on a specific behavior, such as when a person is driving~\cite{fafoutellis2020eco}, or when a particular choice is being made (e.g., using a website or app to select a flight~\cite{sanguinetti2022nudging}). Interfaces using eco-feedback are generally effective at encouraging more sustainable behaviors. For example, displaying information about emissions in vehicle encourages more efficient driving~\cite{fafoutellis2020eco} and interfaces that display the carbon intensity of different options can lead to less carbon intensive choices related to flights~\cite{sanguinetti2022nudging}, ride-sharing and vehicle rental services~\cite{mohanty2023save}, electric vehicle charging schedules~\cite{filipowicz2023promoting}, and food menu items~\cite{beyer2024does}.

However, although people prefer lower carbon intensity options, they do not have a point of reference to compare different metrics of carbon. How much is 1 kg of CO\textsubscript{2}? What does this amount mean for climate change? Mohanty and colleagues~\cite{mohanty2023save} found that although people prioritized ride-sharing and rental vehicle choices that were less carbon intensive, their choices were not influenced by the numeric CO\textsubscript{2} values. This suggests that people use information about carbon as more of a heuristic, preferring lower CO\textsubscript{2} alternatives without knowing whether this lower CO\textsubscript{2} alternative is ``low enough'' compared to some absolute baseline. Interestingly, people in this study preferred being presented with information about \textit{carbon} specifically instead of more relatable alternatives (e.g., equivalent gallons of gasoline) because they saw this information as ``closer to the source''~\cite{mohanty2023save}. Nevertheless, they also expressed a gap in their understanding of CO\textsubscript{2} and the impact of different quantities on climate change, irrespective of the unit used to communicate carbon.

This prior research suggests that although people are motivated to use information about carbon, they need a better grasp of the units used to measure the environmental impact of different behaviors. The following sections explore two different domains---nutrition labeling and physical activity---in which attempts were made to make arbitrary units (calories and step counts) more relevant to people and their choices. We finish with a discussion of the lessons learned from these two domains and how these could apply to design more consistent communication around carbon.

\section{Nutrition labeling}
Rising levels of obesity have been a consistent source of concern for U.S. health officials. In an effort to combat this trend, the U.S. house of representatives introduced the Nutritional Labeling and Education Act of 1990 (NLEA~\cite{NLEA_1990}), which aimed to provide Americans with clear information about the nutrition found in different food items. These labels provide consistent and standardized information about important nutritional elements in food (e.g., calories, sodium, fats, sugars). Although the content of nutrition labels has evolved over the years (e.g., adding trans fats to labels), and additional labeling techniques have also been explored (e.g., ``traffic light'' labels~\cite{traffic}), the NLEA has provided U.S. consumers with consistent and easily accessible nutrition information across all food items for over three decades. 

In addition to labeling the nutrition of household food items, legislation also sought to standardize the communication of nutritional information in restaurant settings. Early policies implemented in cities such as Philadelphia and New York City --- and later nationwide legislation like Section 4205 of the Patient Protection and Affordable Care Act~\cite{ACA_2010} --- required restaurants to display calorie information beside different menu items. The goal of these efforts was to curtail calorie consumption in casual and fast-food restaurant settings, consumption trends that accounted for an increasing proportion of the daily calories consumed by residents~\cite{rosenheck2008fast}.

The goal of legislation like the NLEA and Patient Protection and Affordable Care Act was to use standardized nutritional information displays to increase people's familiarity with abstract nutritional information values (e.g., calories), and help people make more informed nutritional choices. In certain respects, these policies had some success. Since the implementation of the NLEA, there has been a measurable increase in people's awareness and usage of nutritional information when making food purchase choices~\cite{derby2001food, patterson2017unhealthy}. Moreover, frequent and ubiquitous exposure to nutritional labeling has helped increase people's knowledge of different elements found on nutrition labels (calories, sodium, fats)~\cite{drichoutis2005nutrition}. Although nutrition labels are most effectively used by people with high existing knowledge about nutrition, they have also increased familiarity and knowledge about nutrition in people with lower baseline knowledge~\cite{drichoutis2005nutrition, drichoutis2006consumers}.

Similar trends have been observed with the introduction of calorie labels in restaurant settings. Since the implementation of restaurant calorie labels, consumers report being more aware of the calorie content of different restaurant items~\cite{dumanovsky2010consumer} and have more accurate estimates of the calories contained in different food items~\cite{taksler2014calorie, elbel2011consumer}. This suggests that the implementation of wide-spread nutrition information, either through nutrition labels on food items or calories labels in restaurants, increases people's awareness of these different nutritional units, and helps people form better intuitions about how abstract units like calories relate to the food they eat.

Although nutrition labeling tended to increases consumer awareness and knowledge of nutritional information, the impact of this legislation on actual food consumption behavior was less pronounced. A recent meta analysis examining the effect of different standardized nutritional labels suggests a small but consistent effect of nutritional labeling on actual consumption behavior, with labeling accounting for a 6.6\% reduction in food calorie intake~\cite{shangguan2019meta}. However, calorie labeling in restaurants seems less effective at lowering calorie consumption. Although calorie labeling in restaurants increased people's awareness of restaurant food calories, this awareness did not result in lower calorie intake from restaurant foods~\cite{kiszko2014influence, long2015systematic}.

Taken together, this evidence from nutritional labeling suggests that standardized and widespread methods of communicating abstract nutritional information (e.g., calories) can help people become more aware of the nutrition in their food and increase their familiarity and accuracy in estimating nutritional content. Importantly, this labeling can help contextualize abstract units such as calories, giving people better intuitions and reference points across a number of different food items. However, passive information interventions like labeling are likely not sufficient to support real and sustained behavior change and may need to be supported by other intervention mechanisms.

\section{Step Counting}
In addition to changes in nutrition patterns, health research has also sought ways to increase people's levels of daily physical activity. Sedentary lifestyles are associated with numerous health consequences~\cite{park2020sedentary, lavie2019sed} and health practitioners have simple but effective ways to get people engaged in more regular physical exercise. Step counting has become a popular proxy metric of people's ambulatory physical activity~\cite{bassett2017step}. Step counting was popularized by Japanese company Yamax, who had designed a pedometer~\cite{hatano2001pedometer} and introduced the commonly cited target of ``10,000 steps a day'' as part of a marketing campaign. Although research has shown daily step counts under 10,000 can still be beneficial for physical health~\cite{paluch2022daily, kraus2019daily}, step counting goals have become a popular recommendation tool to help people keep track of their daily levels of physical activity. 

Step counting provides a number of advantages as a measure of physical activity. First, regular walking is an effective form of aerobic exercise, and increases in daily step count are consistently associated with better overall health outcomes~\cite{banach2023association, paluch2022daily, hall2020systematic}. Second, as a metric, step counting is objective, unambiguous, and easy to measure, making it a great way for health researchers and consumers to measure and keep track of their physical activity. Last, as a communication tool, step counting is easy for people to understand and implement, providing a simple guideline for programs and broad public outreach efforts aimed at increasing physical activity (e.g., Australia's ``10,000 Step'' program~\cite{vandelanotte2020every}).

Another advantage of step counting is the multitude of tools currently available to help measure and track steps. The growing access to activity trackers has increased people's interest in measuring and tracking their daily activity~\cite{bassett2017step}. Today, step counting is possible on a number of readily accessible platforms, including smartphones and smart watches, offering people many ways to directly track their physical activity. This increased accessibility to activity tracking devices has had an impact on people's actual levels of physical activity. A recent meta analysis found that wearing a devices with activity tracking capabilities increases physical activity by an equivalent of 1800 steps per day~\cite{ferguson2022effectiveness}.

The results from step counting literature suggest that a consistent, easily understandable unit to measure physical activity can help improve peoples physical health. In addition to the simplicity of steps as a metric, the broad access to devices that allow step tracking also greatly facilitate people's awareness of their levels of physical activity~\cite{ferguson2022effectiveness}. Moreover, when the ease of tracking steps is combined with other motivational strategies such as goals--setting (e.g., aim to walk 10,000 steps a day), people show more engagement and better adherence to physical exercise plans~\cite{king2019physical, kraus2019daily}.

\section{Discussion}
This goal of this position paper is to take lessons from nutrition labeling and step counting, and derive recommendations that can apply to sustainable behaviors. As highlighted in the introduction, a challenge with communication about carbon and emissions is that people do not have intuitions for the units associated with climate change (e.g., CO\textsubscript{2}). Our reading of the literature on nutrition labeling and step counting suggests three main recommendations for designers aiming to improve general carbon literacy.

\textbf{1. Display emissions information in a broad range of relevant settings.} A major emphasis of nutritional labeling policy was that the labels be available broadly on all food items. This emphasis ensured that nutrition information could be easily found and compared across a wide range of different products. We argue that the same should apply to information about carbon emissions. Information about emissions should be provided consistently and in a similar format across all relevant domains (e.g., transportation, food, home energy, etc). Providing consistent information across a broad range of domains could help people more accurately represent and compare the emission intensity of different behaviors. If information about emissions is only provided inconsistently, and only for a subset of domains, people may fall into the trends observed in previous research, where they have difficulty understanding the relative impact of different behaviors and risk focusing their efforts on relatively low-impact behaviors~\cite{attari2010public, wynes2018measuring, nielsen2021psychology}.

\textbf{2. Use a single relevant and consistent metric of emissions.} Any metric that is used to communicate information about carbon emissions should be relevant and consistent across domains. Step counting provided people with a clear and understandable metric with which to measure and track their physical activity. Calories provide clear information about the energy content contained in food, a property of food that is directly linked with obesity. We argue that \textbf{CO\textsubscript{2}} or \textbf{CO\textsubscript{2} equivalent} should be used as a consistent unit to communicate emissions. Although a lot of research has focus on ways of communicating emissions through more relatable forms of equivalences (e.g., gallons of gasoline), we argue that using too many disparate metrics may confuse rather than help people compare the impact of different behaviors in different domains. Moreover, research shows that although people do not fully understand what CO\textsubscript{2} represents, communicating CO\textsubscript{2} through eco-feedback interfaces is equal or more effective than different equivalencies at encouraging sustainable choices~\cite{mohanty2023save}. People also reported appreciating a metric that is directly related to climate change~\cite{mohanty2023save}. If information about CO\textsubscript{2} is presented in a broad set of contexts, as outlined in our first recommendation, people may develop better intuitions for the relative carbon intensity of different behaviors, similar to the increase in people's precision at estimating the calorie content of different foods~\cite{taksler2014calorie, elbel2011consumer}.
    
\textbf{3. Combine information interventions with other behavioral change techniques.} Better carbon information displays will likely not be sufficient to fully encourage more sustainable behaviors. A clear result from the nutrition labeling literature is that although labeling increased awareness and knowledge about nutritional information, it had little impact on actual consumption behavior~\cite{shangguan2019meta, kiszko2014influence, long2015systematic}. Similarly, a recent meta-analysis found that information interventions rank among the worst performing interventions to increase more environmentally sustainable behaviors~\cite{bergquist2023field}. However, information interventions can be effective when combined with other intervention types. For example, \emph{goal--setting} can provide a powerful and effective means to increase people's interest in information and promote behavior change~\cite{bergquist2023field}. In the nutrition literature, people who set calorie reduction goals and track their progress see significant reductions in their overall caloric intake~\cite{oh2023inducing}. Similarly, people who set and track goals related to increasing step count also increase their overall physical activity~\cite{king2019physical, kraus2019daily}. If displays or eco-feedback tools incorporate features that allow people to set and track emission reduction goals, these may help people engage more readily with information about emissions and motivate them to change their behavior.

Our recommendations provide some guidelines that designers can use to help people increase carbon literacy and encourage individuals to adopt more environmentally sustainable behaviors. However, we also acknowledge that individual behavior change alone will not be sufficient to address our current climate crisis. The impact of individuals on carbon emissions are far outweighed by the impact of industries that are outside individuals' direct control~\cite{epa2022}. Researchers like Mann~\cite{mann2021new} and Chater and Loewenstein~\cite{chater2023frame} have made appeals to the scientific community to ensure that efforts towards behavioral change do not mask or reduce pressure from the public on wide-spread industrial changes. With that said, the aim of this position piece is to highlight design recommendations that help improve people's carbon literacy. Our hope is that as people develop more accurate intuitions for metrics like carbon, they will not only be able to better judge what behaviors would best for them to change individually, but also recognize how their impact compares to other sources of emissions and be better advocates for real change.

% The ACM Computing Classification System ---
% \url{https://www.acm.org/publications/class-2012} --- is a set of
% classifiers and concepts that describe the computing
% discipline. Authors can select entries from this classification
% system, via \url{https://dl.acm.org/ccs/ccs.cfm}, and generate the
% commands to be included in the \LaTeX\ source.

% User-defined keywords are a comma-separated list of words and phrases
% of the authors' choosing, providing a more flexible way of describing
% the research being presented.

% CCS concepts and user-defined keywords are required for for all
% articles over two pages in length, and are optional for one- and
% two-page articles (or abstracts).

%\section{Acknowledgments}

\bibliographystyle{ACM-Reference-Format}
\bibliography{main}

%%
%% End document here
\end{document}